\documentclass[twocolumn]{aa} 
\usepackage{natbib}
\usepackage{amsmath}
\usepackage{graphicx}
\usepackage{txfonts}
\usepackage{textcomp}

\makeatletter
\def\@biblabel#1{}
\makeatother
\newcommand{\ms}{m\,s$^{-1}$}

\renewcommand{\cite}{\citealp}

\begin{document}

\title{No evidence of the planet orbiting the extremely metal-poor extragalactic star HIP13044.\thanks{Based on data obtained from the ESO Science Archive Facility.}}

\titlerunning{No evidence of the planet HIP13044\,b}

   \author{M. I. Jones \inst{1,2} 
           \and J. S. Jenkins\inst{2} }

         \institute{Departamento de Astronom\'ia y Astrof\'isica, Pontificia Universidad Cat\'olica de Chile, Av. Vicuña Mackenna 4860, 782-0436 Macul, Santiago, Chile \\\email{mjones@aiuc.puc.cl}
         \and Departamento de Astronom\'ia, Universidad de Chile, Camino El Observatorio 1515, Las Condes, Santiago, Chile}

   \date{}

 
  \abstract
{The recent discovery of three giant planets orbiting the extremely metal-poor stars HIP11952 and HIP13044 have 
challenged theoretical predictions of the core-accretion model. According to this, the metal content of the 
protoplanetary disk from which giant planets are formed is a key ingredient for the early formation of 
planetesimals prior to the runaway accretion of the surrounding gas.} 
{We reanalyzed the original FEROS data that were used to detect the planets to prove or refute their existence, employing our new reduction 
and analysis methods.}
{We applied the cross-correlation technique to FEROS spectra to measure the radial velocity variation of HIP13044 and HIP11952. 
We reached a typical precision of $\sim$ 35 \ms\, for HIP13044 and $\sim$ 25 \ms\, for  HIP11952, 
which is significantly superior to the uncertainties presented previously.}
{We found no evidence of the planet orbiting the metal-poor extragalactic star HIP13044. We show that given our radial velocity 
precision, and considering the large number of radial velocity epochs, the probability for a non-detection of the radial velocity signal 
recently claimed is lower than 10$^{-4}$.  Finally, we also confirm findings 
that the extremely metal-poor star HIP11952 does not contain a system of two gas giant planets.  
These results reaffirm the expectations from the core-accretion model of planet formation.} 
   {}


   \maketitle
%

\section{Introduction}

During the past two decades more than 900 extrasolar planets have been detected (http://exoplanet.eu), with a few 
thousand more new candidates that have still to be confirmed (mainly from the KEPLER mission). 
Although the planetary distribution has revealed a huge diversity of planetary properties, there are 
some strong correlations that give us clues about the formation scenarios and the dynamical evolution of these types 
of systems. 
In particular, one of the most important observational results is the planet-metallicity connection. This correlation shows a 
strong increase in the occurrence of giant planets toward more metal-rich stars (Gonzalez \cite{GON97}; Santos et al. 
\cite{SAN01}; Fischer \& Valenti \cite{FIS05}). Moreover, only a few planets have been found around stars with [Fe/H] < 
-0.5. 
This observational result has been used in favor of the core-accretion model of planet formation, where the abundance of metals in the 
protoplanetary disk is an important facet for the growth of cores in the disk, prior to the runaway gas accretion phase (e.g. Ida \& Lin 
\cite{IDA04}; Mordasini et al. \cite{MOR12}). 
The core-accretion model is also supported in the low-mass regime, since there appears to be an under abundance of 
the lowest-mass rocky planets orbiting the most metal-rich stars (Jenkins et al. \cite{JEN13a}), arguing for the rapid growth of planetesimals 
towards the critical core-mass limit. \newline \indent
Recently, Setiawan et al. (\cite{SET10}; hereafter S10) and Setiawan et al. (\cite{SET12}; hereafter S12) announced the detection of two 
planetary systems around the extremely metal-poor stars HIP13044 and HIP11952, respectively. The metallicities of these stars are both at the 
level of [Fe/H] $\sim$ -2~dex, which challenges one of the basic ingredients of the core-accretion model. The discovery of the planet orbiting 
HIP\,13044 conjures up even more fascination, because this star is thought to 
be on the horizontal branch. Additionally, according to its kinematic motion, HIP13044 is part of the Helmi Stream, meaning 
that it originated in a satellite galaxy of the Milky Way (Helmi et al. \cite{HEL99}; Chiba et al. \cite{CHI00}).  
Therefore, this would be the first extragalactic planet ever discovered, highlighting the universality of planet formation.  \newline \indent
In this paper we report the results from a reanalysis of the original S10 and S12 FEROS datasets that were used to discover these planetary systems, along with 
additional observations, using a new method of radial velocity (RV) measurements that we have developed.  Our results indicate that there is no evidence for these two planetary systems. Moreover, in the latter case, we confirm the results recently published by Desidera et al. 
(\cite{DES13}) and Müller et al.  (\cite{MUE13}), who have shown that there is no indication of giant planets orbiting HIP11952.


\section{Observations and data reduction}

We used 46 spectra of HIP13044 that were retrieved from the ESO archive. 
All of the data were taken with FEROS (Kaufer et al. \cite{KAU99}), between the end of 2009 and the beginning of 2011.
The exposure times of the spectra were mostly 900 seconds, with a few of them with an exposure time of 1200 seconds. The typical 
signal-to-noise ratio (S/N) of the spectra is $\sim$ 80-100 at $\sim$ 5500 \AA. 
We used the FEROS Data Reduction System  (DRS) to reduce the raw data. The FEROS pipeline
performs a bias subtraction, flat-field normalization, order-tracing and orders extraction.
The wavelength solution for every order was computed using two calibration lamps taken either
during the afternoon prior to the observations, or during the morning, after the end of the 
observations. The wavelength solution leads to a typical RMS of 0.005 \AA\, from $\sim$ 900 emission lines. \newline \indent
We also retrieved 72 FEROS spectra of the star HIP11952 from the ESO Archive. 
The exposure times and typical S/N values are similar to those of the HIP13044 dataset. The data reduction was performed in the same
fashion as for HIP13044.
\begin{figure*}[t!]
\centering
\includegraphics[scale=0.5,angle=270]{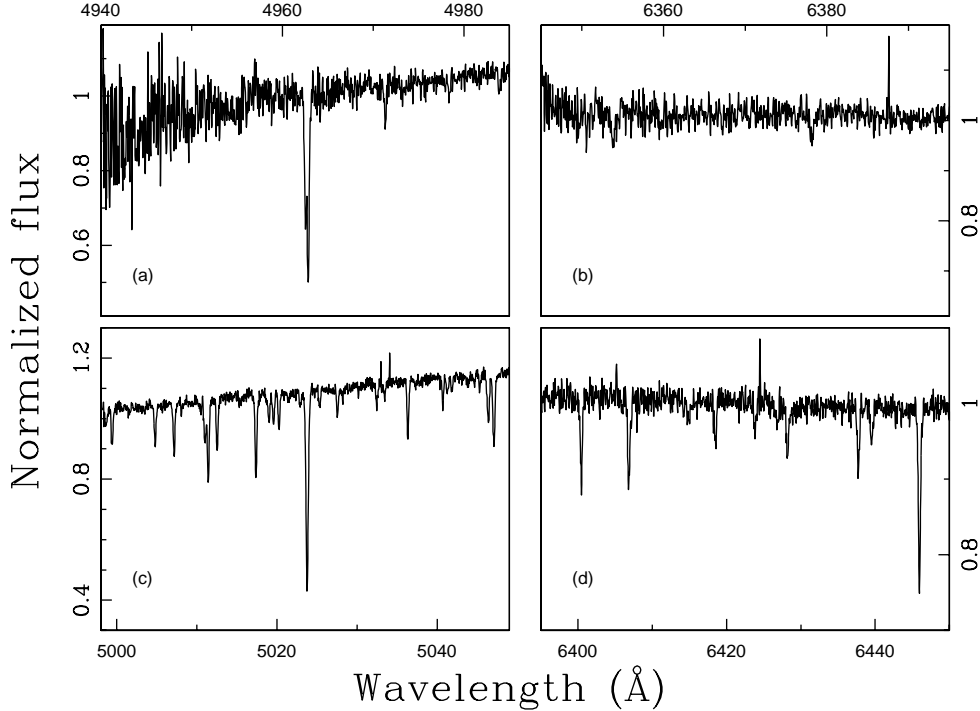}
\caption{Four typical chunks taken from one of the FEROS spectra of HIP13044, showing chunks containing limited RV information (a and b) and chunks relatively rich in RV information (c and d).\label{good_bad_chunks}}
\end{figure*}

\subsection{Radial velocity calculations}

The radial velocities were measured using a similar procedure as the one described in Jones et al. (\cite{JON13}), which has been tested
for G and K dwarfs and giant stars. Based on this method, Jones et al. (\cite{JON13}) showed that it is possible to reach a long-term 
precision of $\sim$ 3 \ms\, using FEROS data. The main steps of this method are described below.\newline \indent
First, we computed the cross-correlation function (Tonry \& Davis \cite{TON79}) between a template and the FEROS spectra.
In this case, the templates correspond to one of the high S/N observations of HIP13044 and HIP11952, instead of 
a numerical mask (e.g. Baranne et al. \cite{BAR96}; Pepe et al. \cite{PEP02}), even though numerical masks have 
proven their ability to generate highly precise RV measurements that can detect planets across a wide range of masses when 
coupled with highly stable instrumentation (e.g. Jenkins et al. \cite{JEN09}; Mayor et al. \cite{MAY09}; Jenkins et al. 
\cite{JEN13b}.) 
We applied this method to 100 chunks of $\sim$ 50 \AA\, width. These chunks were selected from 
25 different orders, covering the region between $\sim$ 4000 and 6500 \AA. 
We then applied  an iterative rejection method that removes every chunk velocity lying 2.5 $\sigma$ away from the mean velocity.
During this process many chunk velocities were rejected, mainly because of the lack of absorption lines in that chunk,
low S/N (especially in the blue edge of each order), and because of telluric lines (at $\sim$ 6000-6500 \AA).
Figure \ref{good_bad_chunks} shows four different chunks at different wavelengths. As can be seen, chunk (a) and (b) are 
noise dominated, and therefore do not contain enough information to measure reliable RVs via the cross-correlation method, whereas 
(c) and (d) do present several absorption lines, hence leading to reliable velocities.
The stellar radial velocity was obtained from the mean in the chunk velocities that were not excluded in the previous step. 
The error bars correspond to the error in the mean. In a similar way, we measured the velocity drift
by cross-correlating the simultaneous lamp (from the sky fiber) with one of the lamps that was used to build the night
wavelength calibration (from the sky fiber as well).  
The resulting RV shift was then subtracted from the measured stellar velocity. 
Finally, we applied the barycentric correction to the measured velocities, using the central time of the observation and the actual coordinates 
of the star, instead of the values recorded in the header (see details in $\S$ 3). 
\begin{figure*}[t!]
\centering
\includegraphics[width=15cm,height=12cm]{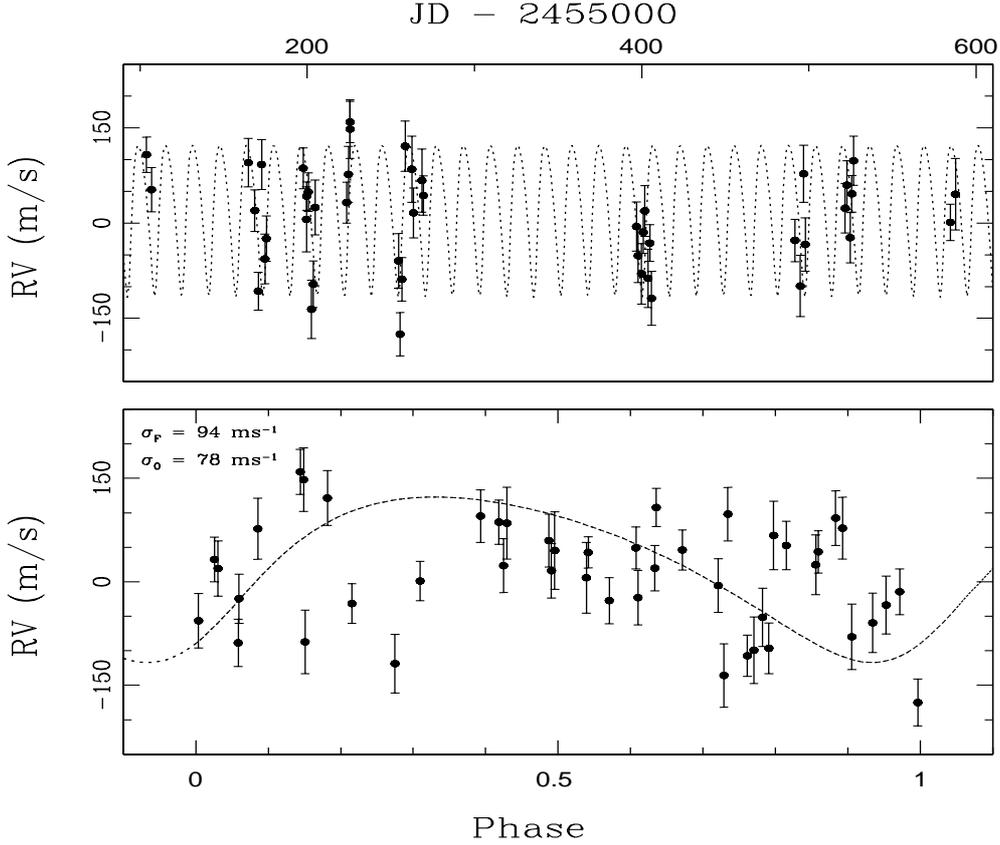}
\caption{Upper Panel: 46 radial velocity measurements of HIP13044 (black dots). The dotted line corresponds to the RV curve predicted by S10.
Lower Panel: Phase folded RVs to the orbital period predicted by S10. The RMS of the fit is 94 \ms, whereas the standard
deviation around zero is 78 \ms. 
\label{HIP13044_RVs}}
\end{figure*}
\begin{figure}
\centering
\includegraphics[width=9cm,height=6cm]{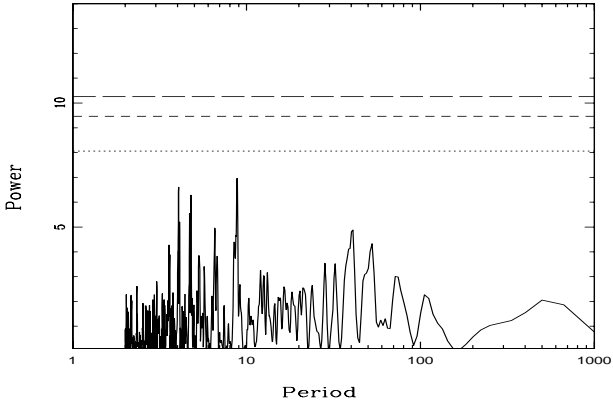}
\caption{LS periodogram of our radial velocities of HIP13044. The three horizontal lines from bottom to top, correspond
to a FAP of 0.01, 0.001 and 0.0001, respectively.
\label{HIP13044_jones_per}}
\end{figure}
Figure \ref{HIP13044_RVs} (upper panel) shows the RV variations computed 
for HIP13044 using our method, which are listed in Table \ref{tab_FEROS_RVs}.
The dotted line corresponds to the RV curve predicted by S10.
The corresponding phase-folded diagram is also plotted (lower panel). The RMS to the S10 solution is 94 \ms. 
Our error bars have typical values of $\sim$ 30-40 \ms. 
The observed standard deviation of our RV measurements is 78 \ms, which is larger than our instrumental errors. 
This probably can be explained by stellar jitter. In fact, HIP13044 
was classified as a horizontal branch star by S10, which are known to exhibit a high level of variability, mainly due to 
stellar oscillations (e.g. Setiawan et al. \cite{SET04}; Hekker et al. \cite{HEK06}). 
The line profile asymmetry analysis presented in S10 (Figure S2) reveals a scatter at the $\sim$ 100 \ms\, level, 
which agrees with our RV measurements.
\begin{table}
\centering
\caption{FEROS RV measurements of HIP13044 \label{tab_FEROS_RVs}}
\begin{tabular}{ccc}
\hline\hline
JD\,-     & RV     & error  \\
2455000   &  (\ms) & (\ms)  \\
\hline
 103.8958 &  107.7 &27.8    \\
 106.8039 &   52.6 &34.9    \\
 164.7661 &   95.3 &38.2    \\
 168.6638 &   19.7 &32.7    \\
 170.7306 & -107.3 &29.8    \\
 172.7060 &   92.3 &39.6    \\
 174.6620 &  -56.5 &39.3    \\
 175.5667 &  -24.5 &35.7    \\
 197.5781 &   86.5 &32.1    \\
 199.5313 &    5.9 &51.2    \\
 199.5742 &   42.5 &22.5    \\
 200.6403 &   49.0 &30.4    \\
 202.6077 & -135.8 &46.0    \\
 203.6124 &  -96.4 &36.8    \\
 204.6604 &   24.6 &43.3    \\
 223.6162 &   32.3 &32.3    \\
 224.5880 &   76.8 &44.2    \\
 225.5371 &  159.3 &32.4    \\
 225.6111 &  148.1 &46.1    \\
 254.5321 &  -59.6 &43.1    \\
 255.5427 & -175.1 &34.0    \\
 256.5481 &  -88.5 &34.2    \\
 258.5431 &  121.3 &39.8    \\
 262.5545 &   84.9 &52.1    \\
 263.5521 &   16.1 &39.5    \\
 268.5183 &   67.1 &49.7    \\
 269.5178 &   43.4 &30.8    \\
 396.8810 &   -5.5 &38.9    \\
 397.8721 &  -51.5 &42.3    \\
 399.8659 &  -79.8 &47.6    \\
 400.9363 &  -14.5 &33.3    \\
 401.8983 &   19.2 &40.0    \\
 403.8446 &  -87.2 &46.0    \\
 404.8909 &  -31.6 &29.0    \\
 405.8531 & -118.7 &42.6    \\
 491.6448 &  -27.3 &33.4    \\
 494.8776 &  -99.2 &48.4    \\
 496.8584 &   77.7 &44.9    \\
 497.8354 &  -33.7 &41.9    \\
 521.6808 &   23.4 &39.0    \\
 522.6921 &   59.8 &38.4    \\
 524.6873 &  -23.0 &39.5    \\
 525.6789 &   46.1 &28.9    \\
 526.6963 &   98.1 &38.7    \\
 584.6150 &    1.1 &28.7    \\
 587.6262 &   45.4 &56.3    \\
\hline
\hline\hline
\end{tabular}
\end{table}
We computed a Lomb-Scargle (LS) periodogram (Scargle \cite{SCA82}) of our RV measurements, which is plotted in Figure \ref{HIP13044_jones_per}.
The three horizontal lines from bottom to top correspond to false alarm
probabilities of 0.01, 0.001 and 0.0001, respectively. As can be seen, there is no significant peak, in the range
between 2 and 1000 days. This result is in stark contrast to the detection of a 16.2-day RV signal found by S10. \newline \indent
To test the reliability of our results, we generated a synthetic dataset by computing
the expected RV value at each observational epoch, using the orbital parameters derived by S10. 
We added to these velocities Gaussian-distributed noise with $\sigma$ equal to the observed standard deviation 
of our RV measurements (78 \ms). Figure \ref{synth_vels} shows the synthetic radial velocities (upper
panel) and the corresponding LS periodogram (lower panel). It can be seen that even including Gaussian 
errors as large as 78 \ms, it is possible to recover the original signal, since the highest peak in the 
periodogram corresponds exactly to the 16.2 days proposed by S10, and has a false alarm probability (FAP) of $\sim$ 10$^{-4}$.
This means that given the large number of data points (a total of 46 compared to only 36 measured by S10), we should be able to  
detect the 16.2-day RV signal claimed by S10, which is obviously not the case. This result forces us to conclude that the RV signal of
16.2 days claimed by S10 is not a real Doppler signal, hence there is no planet orbiting the extragalactic star HIP13044
that has properties in agreement to those announced in S10. 

\subsection{HARPS RVs}

\begin{figure}
\centering
\includegraphics[width=9cm,height=10cm,angle=270]{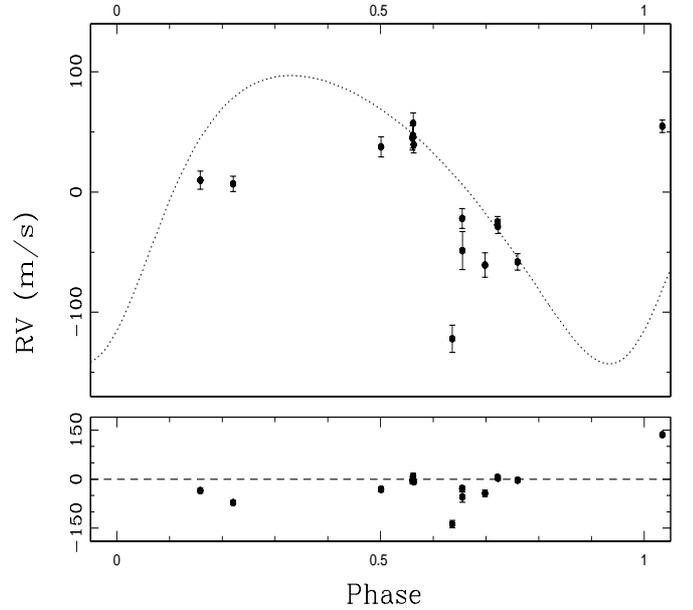}
\caption{HARPS radial velocities measured for HIP13044 (black dots). The dashed line corresponds to the RV signal
predicted by S10. \label{HIP13044_harps}}
\end{figure}

As a secondary test of both the scatter in the RVs for HIP13044 and the reality, or lack thereof, of the proposed planetary system, we 
measured 15 velocities using HARPS.  We observed the star over 9 different nights, during four observing runs.
We used the HARPS DRS to reduce the raw spectra and compute the wavelength solutions, but the procedure of computing the RVs could not 
be used on this star, because no useable order-by-order cross-correlation functions could be generated.  
Therefore, we decided to use the HARPS-TERRA code 
(Anglada-Escude \& Butler \cite{ANGBUT12}) to compute the RVs, which has been used to discover low-mass 
planets (e.g. Anglada-Escude et al. \cite{ANG12}). 
We obtained error bars between 5 \ms and 15 \ms. 
Figure \ref{HIP13044_harps} (upper) shows the resulting RVs (corrected by a zero-point offset of 6.9 \ms). 
The RV values are also listed in Table \ref{tab_HARPS_RVs}. 
The dashed line in Figure \ref{HIP13044_harps} corresponds to the phase-folded signal predicted by S10. The lower panel shows the 
residuals to the S10 solution.
It can be seen the the HARPS velocities and the S10 solution do not seem to be compatible. In particular there are two RV epochs that 
lie $\sim$ 140 \ms\, away from the predicted curve. Moreover, the RMS from a flat curve is $\sim$ 51 \ms, which is smaller than 
the $\sim$ 58 \ms scatter around the S10 solution.
Based on these results, we conclude that the HARPS data also argue against the reality of the planet predicted by S10. 
However, since the number of HARPS observations is very limited, especially because several spectra were taken within the same night 
or in consecutive nights, it is desirable to obtain new HARPS velocities in the future to confirm this result.   

%
\begin{figure}
\centering
\includegraphics[width=7cm,height=8cm,angle=270]{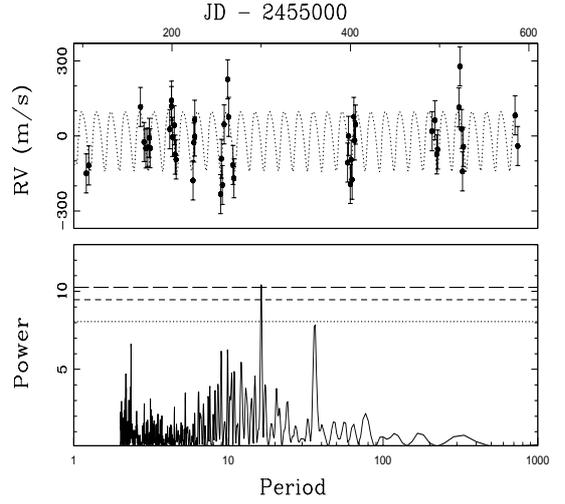}
\caption{Same as Figure \ref{HIP13044_RVs}, but this time using a synthetic dataset, computed from the predicted RV amplitude at
each epoch plus Gaussian noise with a standard deviation equal to 78 \ms.
\label{synth_vels}}
\end{figure}

\section{Possible explanation of the RV discrepancy \label{sec_discrepancies}}

\begin{figure}
\centering
\includegraphics[width=7cm,height=8cm,angle=270]{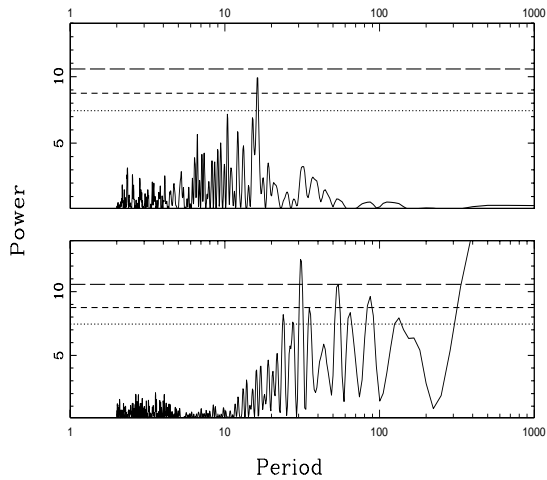}
\caption{LS periodogram of the S10 radial velocities (upper panel) and the sampling (lower panel). 
The three horizontal lines from bottom to top, correspond to a FAP of 0.01, 0.001 and 0.0001, respectively.\label{set_per}}
\end{figure}

\begin{figure}
\centering
\includegraphics[width=7cm,height=8cm,angle=270]{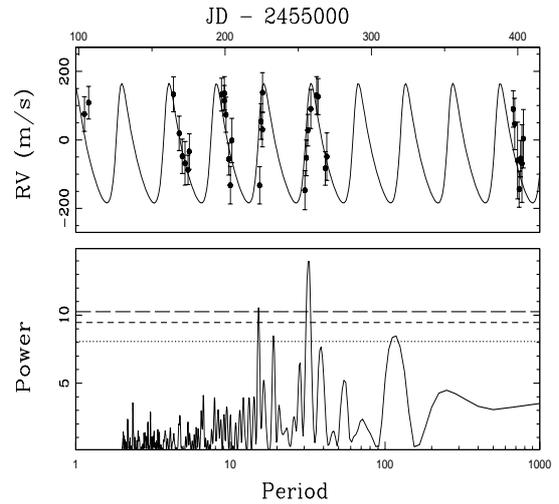}
\caption{Upper panel: Keplerian fit (solid line) of a 32.3-day orbit to the S10 data (black dots), excluding the last two RV 
epochs. The RMS to the fit is 55 \ms. Lower panel: LS periodogram of the S10 data shown in the upper panel. The two highest peaks
correspond to 15.2 days and 32.3 days. The three horizontal lines correspond from bottom to top to FAP of 0.01, 
0.001 and 0.0001 respectively.  \label{32days_fit}}
\end{figure}

\begin{figure}
\centering
\includegraphics[width=8cm,height=9cm,angle=270]{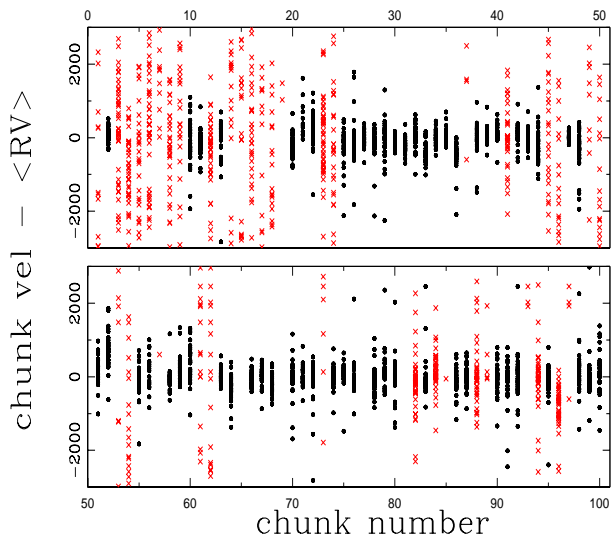}
\caption{Chunk velocities minus the median RV measured at different epochs. The red crosses show the chunks that lead
to a large scatter. The black dots represent chunks that produce a lower dispersion in the radial velocities at 
different epochs, hence those that are included in the analysis.\label{chunks_hip13044}}
\end{figure}

Although the reason for the discrepancy between S10 RV measurements and our velocities is not completely clear, 
we have some ideas of what might have caused the erroneous detection of the periodic RV signal in the S10 data. \newline \indent
First, we noted that the RV signal of 16.2 days claimed by S10 might be caused by the unfortunate combination
of poor signal-to-noise ratio and the window function (sampling). Figure \ref{set_per} shows the LS periodogram of the
RV variations (upper panel) and the sampling (lower panel). In the upper panel we can clearly distinguish the 16.2-day 
peak, but in the lower panel, there is a peak at 31.4 days, which is very close to twice the 16.2-day period. 
Moreover, we noted that after removing the last two RV epochs in the S10 dataset it is possible to fit a 32.3-day orbit, which leads to 
almost the same RMS as the original fit (see upper panel of Figure \ref{32days_fit}). 
Furthermore, we found that the new periodogram of the data shows a very strong peak at 32.3 days (lower panel in Figure \ref{32days_fit}). 
These results strongly suggest that the RV signal claimed by S10 is related to the sampling instead of to a genuine Doppler signal in the data.
\newline  \indent
A second problem that we found in the S10 data is related to the barycentric correction. S10 mentioned that they used the FEROS DRS 
to produce 39 one-dimensional spectra, that were shifted to the rest frame after applying the barycentric velocity 
correction. However, this approach is not correct, since the FEROS-DRS barycentric correction introduces large systematic 
errors (see M\"{u}ller et al. \cite{MUE13}).
Additionally, the julian dates given by S10 correspond to the beginning of the
observations instead of to the central time of the observation. For this specific dataset, this corresponds to a difference in time of $\sim$ 
8-10 minutes, leading to systematic errors in the barycentric velocity correction as large as $\sim$ 15 \ms. These two 
errors, when combined, can produce systematic shifts in radial velocity exceeding 20 \ms. \newline \indent
Finally, we note that since there are large portions in most of the orders where there is a lack of absorption lines, 
the order-by-order cross-correlation can lead to inaccurate results. While S10 obtained typical error bars of $\sim$ 50-80 
\ms, we obtained uncertainties at the $\sim$ 30-40 \ms\, level. 
The reason why we obtained smaller error bars is because we computed the cross-correlation function in regions of 
$\sim$ 50 \AA, rejecting chunks that lead to very deviant velocities. 
Figure \ref{chunks_hip13044} shows the difference between each chunk velocity and the median velocity from all chunks at 
each epoch (hence at each chunk number there are 46 velocities in the y-axis direction). 
As can be seen, there are many chunks that produce deviant velocities at different epochs, which are rejected by our 
code (red crosses). On the other hand, chunks that lead to tighter velocities at different epochs are 
included in the final RV analysis (mostly corresponding to the black dots).
It is worth mentioning that the chunks leading to constant velocities in time (black dots in Figure \ref{chunks_hip13044}) do 
not produce RVs close to zero, but they lead to radial velocities as high as 
$\sim$ 30 k\ms. Only after the barycentric velocity correction is applied, and the nightly drift is substracted, do they 
become "flat". 
It is evident from Figure \ref{chunks_hip13044} that including a complete order (compounded by four consecutive chunks) in the 
cross-correlation function produces a larger scatter in the final radial velocities. 

\section{HIP11952}

\begin{figure}
\centering
\includegraphics[width=7cm,height=9cm,angle=270]{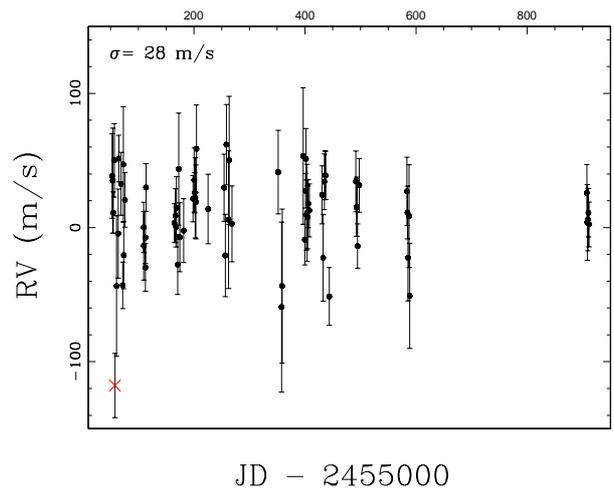}
\caption{RV measurements of HIP11952. The red cross corresponds to a deviant point that was excluded   
from the analysis. The error bars are typically $\sim$ 20-30 \ms. The scatter around zero is 28 \ms. \label{HIP11952}}
\end{figure}

As a part of the same RV program, S12 announced the detection of a planetary system around the 
extremely metal-poor ([Fe/H]=-1.9) star HIP11952. As for HIP13044, this discovery was very controversial
since according to the core-accretion model, giant planets are not expected to be formed around such a metal-poor star, 
as we discussed above.
However, based on much higher quality data taken with HARPS-N, Desidera et al. (\cite{DES13}) recently showed that
there is no planetary system around HIP11952 and that the radial velocities are flat, showing a small scatter at the 7 \ms\, level, 
which has also been confirmed by M\"{u}ller et al. (\cite{MUE13}). \newline \indent
\begin{table}
\centering
\caption{HARPS RV measurements of HIP13044 \label{tab_HARPS_RVs}}
\begin{tabular}{ccc}
\hline\hline
JD\,-     & RV     & error  \\
2455000   &  (\ms) & (\ms)  \\
\hline
 1448.9294 & -128.9 &  11.3 \\
 1449.9362 &  -67.6 &  10.1 \\
 1450.9352 &  -65.0 &   6.8 \\
 1462.9414 &   30.8 &   8.3 \\
 1463.9067 &   38.2 &  10.1 \\
 1463.9171 &   40.3 &  10.5 \\
 1463.9299 &   50.4 &  8.6 \\
 1463.9445 &   32.6 &   6.9 \\
 1562.6326 &  -28.8 &   8.2 \\
 1562.6393 &  -55.6 &  15.8 \\
 1563.7164 &  -31.9 &   4.7 \\
 1563.7306 &  -35.5 &   5.9 \\
 1633.5839 &   47.9 &   5.3 \\
 1635.5868 &    3.0 &   7.6 \\
 1636.5963 &    0.0 &   6.3 \\
\hline\hline
\end{tabular}
\end{table}
To test this result and the reliability of our method, we also computed the radial velocities using 72 FEROS spectra of HIP11952 
in the same manner as explained above. Figure \ref{HIP11952} shows the 
resulting RVs. The red cross corresponds to a deviant data point. As can be seen, the RV curve of HIP11952 is flat, showing a
scatter of only 28 \ms, which can be explained solely by instrumental errors, since the mean value of the error bars is 27 \ms,
nearly twice as good as the S12 uncertainties on the same data.
Clearly, there is no indication of the two signals claimed by S12, both of them with predicted amplitudes above 100 \ms.
This result proves that our method leads to consistent radial velocity measurements, even for extremely metal-deficient stars, 
reinforcing our results on HIP13044, and it also confirms that HIP11952 does not host a planetary system with the characteristics 
claimed by S12.

\section{Discussion}

Based on the radial velocity analysis of 46 FEROS spectra, we found no evidence for a giant planet around HIP13044, in contrast to the 
results announced by S10, who claimed the detection of a periodic RV signal that was interpreted to 
be caused by a giant planet with an orbital period of 16.2 days.  We also confirm the flat RV time series for another 
extremely metal-poor star HIP11952, which was claimed in S12 to have a system of two giant planets. These results 
again confirm the core-accretion 
scenario for planet formation, adding weight to the argument that it is difficult to form gas giant planets in dust-depleted protoplanetary-planetary disks. \newline \indent 
HIP13044 was classified as a horizontal branch star by S10, meaning that HIP13044\,$b$ could have been a planet that had 
survived the common-envelope phase, since the radii of stars at the end of the red giant phase are much larger than the 
orbital distance of the proposed planet. In fact, no planets have yet been detected orbiting interior to $\sim$ 0.5 AU around giant stars. 
Our results show that there is no need to develop complex models to explain the existence of this planet, and how it could have survived this 
disastrous phase of stellar evolution (Bear et al. \cite{BEA11}; Passy et al. \cite{PAS12}). \newline \indent
Based on the kinematic properties of HIP13044, S10 highlighted that this star has an extragalactic origin, meaning the planet 
orbiting this star would have been the first planet known to originate from beyond the Milky Way galaxy. 
Our results state however, that there are still no planets around extremely metal-poor stars and that all of them have been
formed around stars originating within our Galaxy. 


\begin{acknowledgements}
M.J. acknowledges financial support from Fondecyt grant \#1120299 and  
ALMA-Conicyt grant \#31080027. J.J. acknowledges funding by Fondecyt through grant 3110004, 
the GEMINI-CONICYT FUND and from the Comit\'e Mixto ESO-GOBIERNO DE CHILE.
We also acknowledge support from Basal PFB-06 (CATA). 
\end{acknowledgements}

\end{document}